\documentclass{article}%
\usepackage{amsmath}%
\usepackage{amsfonts}%
\usepackage{amssymb}%
\usepackage{graphicx}

\begin{document}

\title{Representaciones integrales para el campo electromagnetico en medios chirales}
\author{Vladislav Kravchenko, Hector Oviedo\\Depto. de Telecomunicaciones,\\Escuela Superior de Ingenier\'{\i}a Mec\'{a}nica y El\'{e}ctrica,\\Instituto Polit\'{e}cnico Nacional,\\C.P.07738, D.F., MEXICO\\e-mail: vkravche@maya.esimez.ipn.mx oviemac@mail.internet.com.mx}
\maketitle

\begin{abstract}
Se presentan las nuevas representaciones integrales para las soluciones de las
ecuaciones de Maxwell en medios chirales. Las representaciones integrales se
obtienen con la ayuda\ de los metodos del analisis
cuaternionico\ desarrollados recientemente.

\end{abstract}

\bigskip

\section{Introducci\'{o}n}

En los \'{u}ltimos a\~{n}os la construcci\'{o}n de elementos compuestos de
material chiral ha adquirido gran importancia. En muchos campos de la
f\'{\i}sica, en particular en las telecomunicaciones se utilizan materiales
chirales, en especial en la fabricaci\'{o}n de antenas. Los materiales
\ chirales se encuentran en muchos productos en forma natural como
mol\'{e}culas \'{o}pticas activas, sin embargo, la aplicaci\'{o}n masiva ha
llevado a su producci\'{o}n artificial, habiendo actualmente muchos
laboratorios dedicados a producir material chiral.

En este trabajo se presentan las nuevas representaciones integrales para las
soluciones de las ecuaciones de Maxwell y se resuelve el problema de
extendibilidad del campo electromagn\'{e}tico en medios chirales. Los
resultados se obtienen con la ayuda\ de los m\'{e}todos del an\'{a}lisis
cuaterni\'{o}nico\ desarrollados recientemente.

\bigskip

\section{\bigskip Algunos elementos del an\'{a}lisis cuaterni\'{o}nico}

Es necesario presentar algunos elementos del an\'{a}lisis cuaterni\'{o}nico
que se usar\'{a}n en este desarrollo.Consid\'{e}rese el conjunto de
cuaterniones complejos como $\mathbb{H}\left(  \mathbb{C}\right)  $. Un
cuaterni\'{o}n $q\in\mathbb{H}\left(  \mathbb{C}\right)  $ se representa por
$q=\sum\limits_{k=0}^{3}q_{k}i_{k}$ donde $\{q_{k}\}\subset\mathbb{C}$,
$i_{0}=1 $ e $i_{k}$, $k=1,2,3$ son las unidades cuaterni\'{o}nicas
imaginarias est\'{a}ndard con: $i_{k}^{2}=-1$, $k=1,2,3$; $i_{1}i_{2}%
=-i_{2}i_{1}=i_{3}$, $i_{2}i_{3}=-i_{3}i_{2}=i_{1}$, $i_{3}i_{1}=-i_{1}%
i_{3}=i_{2}$. Por definici\'{o}n, la unidad imaginaria compleja $i$ conmuta
con $i_{k}$, $k=0,1,2,3$. Se utilizar\'{a} tambi\'{e}n la representaci\'{o}n
vectorial de los cuaterniones complejos, esto es, cualquier $q\in
\mathbb{H}\left(  \mathbb{C}\right)  $ se puede representar en la forma
$q=$Sc$\left(  q\right)  +$Vec$\left(  q\right)  $, donde $\operatorname*{Sc}%
\left(  q\right)  :=q_{0}$ y $\operatorname*{Vec}\left(  q\right)
:=\sum\limits_{k=1}^{3}q_{k}i_{k}$. Los cuaterniones complejos de la forma
$q=\operatorname*{Vec}\left(  q\right)  $ se dice que son cuaterniones
puramente vectoriales y se representan por $\overrightarrow{q}$. Note que
\'{e}stos se pueden identificar con los vectores en $C^{3}$.

\bigskip

De la misma manera que en los n\'{u}meros complejos, se define el conjugado de
un cuaterni\'{o}n complejo $q$ como $\overline{q}:=$Sc$\left(  q\right)
-$Vec$\left(  q\right)  =q_{0}-\overrightarrow{q}$.

\bigskip

Consid\'{e}rense ahora las funciones valuadas en $\mathbb{H}\left(
\mathbb{C}\right)  $ dadas en alg\'{u}n dominio $\Omega\subset\Re^{3}$ . En
todo lo que sigue vamos a suponer que la frontera $\Gamma$ del dominio
$\Omega$ es una superficie de Liapunov. (vea por ejemplo \cite{Vladimirov}).
Sobre el conjunto $C^{1}\left(  \Omega\text{;}\mathbb{H}\left(  \mathbb{C}%
\right)  \right)  $ se define el conocido operador de Moisil-Theodoresco (vea
por ejemplo \cite{Bit} y \cite{BDS}) por la expresi\'{o}n

\bigskip%
\begin{equation}
D:=i_{1}\frac{\partial}{\partial x_{1}}+i_{2}\frac{\partial}{\partial x_{2}%
}+i_{3}\frac{\partial}{\partial x_{3}} \label{S1}%
\end{equation}

\bigskip Sea la funci\'{o}n $f$ en $C^{1}\left(  \Omega\text{;}\mathbb{H}%
\left(  \mathbb{C}\right)  \right)  $. La expresi\'{o}n $Df$ \ se puede
reescribir de la siguiente manera:%

\begin{equation}
Df=-\text{div}\overrightarrow{f}+\text{grad}f_{0}+\text{rot}\overrightarrow{f}
\label{S2}%
\end{equation}
donde los operadores diferenciales se definen de la manera usual. Por ejemplo,

\bigskip%

\begin{equation}
\text{grad}f_{0}=i_{1}\frac{\partial}{\partial x_{1}}f_{0}+i_{2}%
\frac{\partial}{\partial x_{2}}f_{0}+i_{3}\frac{\partial}{\partial x_{3}}%
f_{0}\text{.} \label{S3}%
\end{equation}

\bigskip

\bigskip El lado derecho de la ecuaci\'{o}n (\ref{S2}) no tendr\'{\i}a sentido
en c\'{a}lculo vectorial, aqu\'{\i} simplemente significa un cuaterni\'{o}n
completo con su parte escalar y su parte vectorial.

\bigskip%
\begin{equation}
Sc\left(  Df\right)  =-\text{div}\overrightarrow{f},\ \ \ \ Vec\left(
Df\right)  =\text{grad}f_{0}+\text{rot}\overrightarrow{f} \label{S4}%
\end{equation}

Una propiedad importante de $D$ es $D^{2}=-\Delta$ donde $\Delta
=\frac{\partial^{2}}{\partial x_{1}^{2}}+\frac{\partial^{2}}{\partial
x_{2}^{2}}+\frac{\partial^{2}}{\partial x_{3}^{2}}$ es el operador de Laplace.

Consid\'{e}rese el siguiente operador
\begin{equation}
D_{\pm\alpha}:=D\pm\alpha\label{S5}%
\end{equation}

\bigskip Y los operadores \ $T_{\pm\alpha}$ y $K_{\pm\alpha}$ de la siguiente manera:%

\begin{equation}
T_{\pm\alpha}f\left(  x\right)  =\int\limits_{\Omega}\Upsilon_{\pm\alpha
}\left(  x-y\right)  f\left(  y\right)  dy \label{S6}%
\end{equation}%

\begin{equation}
K_{\pm\alpha}f\left(  x\right)  =-\int\limits_{\Gamma}\Upsilon_{\pm\alpha
}\left(  x-y\right)  \overrightarrow{n}\left(  y\right)  f\left(  y\right)
d\Gamma_{y} \label{S7}%
\end{equation}
donde $\Upsilon_{\pm\alpha}$ es la soluci\'{o}n fundamental de $D_{\pm\alpha}$
dada por

\smallskip%
\begin{align}
\Upsilon_{\pm\alpha}\left(  x\right)   &  =\Theta_{\alpha}\left(  x\right)
\left(  \pm\alpha+\frac{x}{\left|  x\right|  ^{2}}+i\alpha\frac{x}{\left|
x\right|  }\right) \label{S8}\\
&  =-\text{grad}\Theta_{\alpha}\left(  x\right)  \pm\alpha\Theta_{\alpha
}\left(  x\right)
\end{align}

\bigskip

Aqu\'{\i} \bigskip$\Theta_{\alpha}$ es la soluci\'{o}n fundamental del
operador de Helmholtz \ $\Theta_{\alpha}(x)=-\frac{e^{i\alpha\left|  x\right|
}}{4\pi\left|  x\right|  }.$

En este punto es importante mencionar el teorema (ver pag. 75 de
\cite{KSbook}) para la f\'{o}rmula cuaterni\'{o}nica de Borel -Pompeiu que dice:

\bigskip

Teorema.- Sea $\alpha$ un cuaterni\'{o}n complejo arbitrario, $\Omega$ un
dominio en $R^{3}$ con la frontera de Liapunov $\Gamma:=\partial\Omega$, y
sean $T_{\alpha}$, $K_{\alpha}$ como se muestra en (\ref{S6}) y (\ref{S7}),
entonces, si $f\in C^{1}(\Omega;\mathbb{H(C)})$ $\cap$ $C(\overline{\Omega
};\mathbb{H(C)})$ se cumple que $\forall x\in\Omega$ se satisface que

\smallskip%
\begin{equation}
(K_{\alpha}+T_{\alpha}D_{\alpha})f\left(  x\right)  =f\left(  x\right)
\label{S9}%
\end{equation}

\section{Los Campos Electromagn\'{e}ticos en Medios Chirales.}

Consid\'{e}rense las ecuaciones de Maxwell\bigskip%

\begin{equation}
\operatorname*{div}\widetilde{E}\left(  x\right)  =\frac{\rho}{\varepsilon},
\label{S10}%
\end{equation}%

\begin{equation}
\operatorname{rot}\widetilde{E}\left(  x\right)  =-i\omega\widetilde{B}\left(
x\right)  , \label{S11}%
\end{equation}%

\begin{equation}
\operatorname{rot}\widetilde{H}\left(  x\right)  =i\omega\widetilde{D}\left(
x\right)  +\widetilde{j}\left(  x\right)  , \label{S12}%
\end{equation}%

\begin{equation}
\operatorname*{div}\widetilde{H}\left(  x\right)  =0 \label{S13}%
\end{equation}
en donde las ecuaciones materiales est\'{a}n dadas por (vea por ejemplo
\cite{LVV})
\begin{equation}
\widetilde{D}=\varepsilon\left(  \widetilde{E}\left(  x\right)  +\beta
\operatorname{rot}\widetilde{E}\left(  x\right)  \right)  \label{S14}%
\end{equation}
\begin{equation}
\widetilde{B}=\mu\left(  \widetilde{H}\left(  x\right)  +\beta
\operatorname{rot}\widetilde{H}\left(  x\right)  \right)  \label{S15}%
\end{equation}
\smallskip el par\'{a}metro $\beta$ es la medida de chiralidad del medio.
Entonces, las ecuaciones (\ref{S11}) y (\ref{S12}) se pueden escribir como%

\begin{equation}
\operatorname{rot}\widetilde{E}\left(  x\right)  =-i\omega\mu\left(
\widetilde{H}\left(  x\right)  +\beta\operatorname{rot}\widetilde{H}\left(
x\right)  \right)  \label{S16}%
\end{equation}%

\begin{equation}
\operatorname{rot}\widetilde{H}\left(  x\right)  =i\omega\varepsilon\left(
\widetilde{E}\left(  x\right)  +\beta\operatorname{rot}\widetilde{E}\left(
x\right)  \right)  +\widetilde{j}\left(  x\right)  \label{S17}%
\end{equation}

Ahora, para simplificar estas ecuaciones, introduzcamos los siguientes
vectores del campo%

\begin{equation}
\widetilde{E}\left(  x\right)  =\sqrt{\mu}\cdot\overrightarrow{E}\left(
x\right)  \label{S18}%
\end{equation}%

\begin{equation}
\widetilde{H}\left(  x\right)  =\sqrt{\varepsilon}\cdot\overrightarrow
{H}\left(  x\right)  \label{S19}%
\end{equation}%

\begin{equation}
\widetilde{j}\left(  x\right)  =\sqrt{\varepsilon}\cdot\overrightarrow
{j}\left(  x\right)  \label{S20}%
\end{equation}
resultando las expresiones\bigskip%

\begin{equation}
\operatorname{rot}\overrightarrow{E}\left(  x\right)  =-ik\left(
\overrightarrow{H}\left(  x\right)  +\beta\operatorname{rot}\overrightarrow
{H}\left(  x\right)  \right)  \label{S21}%
\end{equation}%

\begin{equation}
\operatorname{rot}\overrightarrow{H}\left(  x\right)  =ik\left(
\overrightarrow{E}\left(  x\right)  +\beta\operatorname{rot}\overrightarrow
{E}\left(  x\right)  \right)  +\overrightarrow{j}\left(  x\right)  \label{S22}%
\end{equation}
donde $k=\omega\sqrt{\mu\varepsilon}$ , \ \ $k$ \ es el n\'{u}mero de onda

\bigskip

\section{Las ecuaciones de Maxwell para medios chirales en forma cuaterni\'{o}nica}

Haciendo uso del c\'{a}lculo cuaterni\'{o}nico, las ecuaciones (\ref{S21}) y
(\ref{S22}) se pueden reescribir en tal forma que las funciones vectoriales
que estamos buscando queden en ecuaciones separadas.

\bigskip

Para este prop\'{o}sito, se introducen las siguientes funciones
cuaterni\'{o}nicas puramente vectoriales:%

\begin{equation}
\overrightarrow{\Phi}=\overrightarrow{E}\left(  x\right)  +i\overrightarrow
{H}\left(  x\right)  \label{S23}%
\end{equation}%

\begin{equation}
\overrightarrow{\Psi}=\overrightarrow{E}\left(  x\right)  -i\overrightarrow
{H}\left(  x\right)  \label{S24}%
\end{equation}%

\begin{equation}
\overrightarrow{E}\left(  x\right)  =\frac{1}{2}\left(  \overrightarrow{\Phi
}+\overrightarrow{\Psi}\right)  \left(  x\right)  \label{S25}%
\end{equation}%

\begin{equation}
\overrightarrow{H}\left(  x\right)  =\frac{1}{2i}\left(  \overrightarrow{\Phi
}-\overrightarrow{\Psi}\right)  \left(  x\right)  \label{S26}%
\end{equation}

Al aplicar el operador de Moisil-Theodoresco a la ecuaci\'{o}n (\ref{S23})%

\begin{equation}
D\overrightarrow{\Phi}=D\overrightarrow{E}\left(  x\right)  +iD\overrightarrow
{H}\left(  x\right)  \label{S27}%
\end{equation}%

\begin{equation}
D\overrightarrow{E}\left(  x\right)  =-\operatorname*{div}\overrightarrow
{E}\left(  x\right)  +\operatorname{rot}\overrightarrow{E}\left(  x\right)
=-\frac{\rho}{\varepsilon}+\operatorname{rot}\overrightarrow{E}\left(
x\right)  \label{S28}%
\end{equation}%

\begin{equation}
D\overrightarrow{H}\left(  x\right)  =-\operatorname*{div}\overrightarrow
{H}\left(  x\right)  +\operatorname{rot}\overrightarrow{H}\left(  x\right)
=\operatorname{rot}\overrightarrow{H}\left(  x\right)  \label{S29}%
\end{equation}%

\begin{equation}
D\overrightarrow{\Phi}\left(  x\right)  =-\frac{\rho}{\varepsilon
}+\operatorname{rot}\overrightarrow{E}\left(  x\right)  +i\operatorname{rot}%
\overrightarrow{H}\left(  x\right)  \label{S30}%
\end{equation}

Utilizando las ecuaciones\bigskip(\ref{S21}) y (\ref{S22}) se obtiene%

\begin{equation}
\left(  D+\frac{k}{\left(  1+k\beta\right)  }\text{ }\right)  \overrightarrow
{\Phi}\left(  x\right)  =-\frac{\rho}{\varepsilon}+i\frac{\overrightarrow
{j}\left(  x\right)  }{\left(  1+k\beta\right)  } \label{S39}%
\end{equation}
\smallskip

Los mismos pasos se siguen a la ecuaci\'{o}n (\ref{S24}) para obtener\smallskip%

\begin{equation}
\left(  D-\frac{k}{\left(  1-k\beta\right)  }\right)  \overrightarrow{\Psi
}\left(  x\right)  =-\frac{\rho}{\varepsilon}-i\frac{\overrightarrow{j}\left(
x\right)  }{\left(  1-k\beta\right)  } \label{S40}%
\end{equation}
\smallskip

Al aplicar la divergencia a la ecuaci\'{o}n (\ref{S22}) se obtiene la
ecuaci\'{o}n de continuidad%

\begin{equation}
\frac{\rho}{\varepsilon}=-\frac{1}{ik}\operatorname*{div}\overrightarrow
{j}\left(  x\right)  \label{S44}%
\end{equation}
la que al sustituir en las ecuaciones (\ref{S39}) y (\ref{S40}) da como resultado:\smallskip%

\begin{equation}
\left(  D+\frac{k}{\left(  1+k\beta\right)  }\right)  \overrightarrow{\Phi
}\left(  x\right)  =i\frac{\overrightarrow{j}\left(  x\right)  }{\left(
1+k\beta\right)  }-\frac{i}{k}\operatorname*{div}\overrightarrow{j}\left(
x\right)  \label{S45}%
\end{equation}%

\begin{equation}
\left(  D-\frac{k}{\left(  1-k\beta\right)  }\right)  \overrightarrow{\Psi
}\left(  x\right)  =-i\frac{\overrightarrow{j}\left(  x\right)  }{\left(
1-k\beta\right)  }-\frac{i}{k}\operatorname*{div}\overrightarrow{j}\left(
x\right)  \label{S46}%
\end{equation}

Introduciendo las siguientes notaciones: $\alpha_{1}:=\frac{k}{(1+k\beta)}$,
$\alpha_{2}:=\frac{k}{(1-k\beta)}$ obtenemos estas ecuaciones en la forma:

\bigskip%

\begin{equation}
(D+\alpha_{1})\overrightarrow{\Phi}\left(  x\right)  =\frac{i}{k}[\alpha
_{1}\overrightarrow{j}\left(  x\right)  -\operatorname*{div}\overrightarrow
{j}\left(  x\right)  ]
\end{equation}%

\begin{equation}
(D-\alpha_{2})\overrightarrow{\Psi}\left(  x\right)  =-\frac{i}{k}[\alpha
_{2}\overrightarrow{j}\left(  x\right)  +\operatorname*{div}\overrightarrow
{j}\left(  x\right)  ]
\end{equation}

\bigskip

\section{Representaci\'{o}n integral para las soluciones de las ecuaciones de
Maxwell en medios chirales}

\bigskip

Sustituyendo en (\ref{S9}) los resultados obtenidos%

\begin{equation}
\overrightarrow{\Phi}\left(  x\right)  =\frac{i}{k}T_{\alpha_{1}}[\alpha
_{1}\overrightarrow{j}\left(  x\right)  -\operatorname*{div}\overrightarrow
{j}\left(  x\right)  ]+K_{\alpha_{1}}\overrightarrow{\Phi}\left(  x\right)
\label{S47}%
\end{equation}%

\begin{equation}
\overrightarrow{\Psi}\left(  x\right)  =-\frac{i}{k}T_{-\alpha_{2}}[\alpha
_{2}\overrightarrow{j}\left(  x\right)  +\operatorname*{div}\overrightarrow
{j}\left(  x\right)  ]+K_{-\alpha_{2}}\overrightarrow{\Psi}\left(  x\right)
\label{S48}%
\end{equation}
donde $x$ es un punto arbitrario del dominio $\Omega$.

Ahora, si $T_{\alpha_{1}}=T_{1\text{, \ }}$ $T_{-\alpha_{2}}=T_{2}$,
\ $K_{\alpha_{1}}=K_{1}$, \ $K_{-\alpha_{2}}=K_{2}$, $\Theta_{\alpha_{1}%
}=\Theta_{1}$ y $\Theta_{-\alpha_{2}}=\Theta_{2}$\ \ y usando estos resultados
en las ecuaciones (\ref{S25}) y (\ref{S26}) se obtiene para el campo
el\'{e}ctrico la expresi\'{o}n\smallskip\
\begin{align}
\overrightarrow{E}\left(  x\right)   &  =\frac{1}{2}\{iT_{1}%
[\frac{\overrightarrow{j}\left(  x\right)  }{\left(  1+k\beta\right)
}-\frac{1}{k}\operatorname*{div}\overrightarrow{j}\left(  x\right)
]\nonumber\\
&  +K_{1}[\overrightarrow{E}\left(  x\right)  +i\overrightarrow{H}\left(
x\right)  ]\nonumber\\
&  -iT_{2}[\frac{\overrightarrow{j}\left(  x\right)  }{\left(  1-k\beta
\right)  }+\frac{1}{k}\operatorname*{div}\overrightarrow{j}\left(  x\right)
]\nonumber\\
&  +K_{2}[\overrightarrow{E}\left(  x\right)  -i\overrightarrow{H}\left(
x\right)  ]\}
\end{align}

y para el campo magn\'{e}tico%

\begin{align}
\overrightarrow{H}\left(  x\right)   &  =\frac{1}{2i}\{iT_{1}%
[\frac{\overrightarrow{j}\left(  x\right)  }{\left(  1+k\beta\right)
}-\frac{1}{k}\operatorname*{div}\overrightarrow{j}\left(  x\right)
]\nonumber\\
&  +K_{1}[\overrightarrow{E}\left(  x\right)  +i\overrightarrow{H}\left(
x\right)  ]\nonumber\\
&  +iT_{2}[\frac{\overrightarrow{j}\left(  x\right)  }{\left(  1-k\beta
\right)  }+\frac{1}{k}\operatorname*{div}\overrightarrow{j}\left(  x\right)
]\nonumber\\
&  -K_{2}[\overrightarrow{E}\left(  x\right)  -i\overrightarrow{H}\left(
x\right)  ]\}\nonumber
\end{align}
\smallskip

y m\'{a}s expl\'{\i}citamente\bigskip%
\begin{align}
\overrightarrow{E}\left(  x\right)   &  =\frac{i}{2}\int_{\Omega}%
\{\Upsilon_{\alpha_{1}}\left(  x-y\right)  [\frac{\overrightarrow{j}\left(
y\right)  }{\left(  1+k\beta\right)  }-\frac{1}{k}\operatorname*{div}%
\overrightarrow{j}\left(  y\right)  ]\nonumber\\
&  -\Upsilon_{-\alpha_{2}}\left(  x-y\right)  [\frac{\overrightarrow{j}\left(
y\right)  }{\left(  1-k\beta\right)  }+\frac{1}{k}\operatorname*{div}%
\overrightarrow{j}\left(  y\right)  ]\}dy\nonumber\\
&  -\frac{1}{2}\int_{\Gamma}\{\Upsilon_{\alpha_{1}}\left(  x-y\right)
\overrightarrow{n}\left(  y\right)  [\overrightarrow{E}\left(  y\right)
+i\overrightarrow{H}\left(  y\right)  ]\nonumber\\
&  +\Upsilon_{-\alpha_{2}}\left(  x-y\right)  \overrightarrow{n}\left(
y\right)  [\overrightarrow{E}\left(  y\right)  -i\overrightarrow{H}\left(
y\right)  ]\}d\Gamma
\end{align}

\bigskip%
\begin{align}
\overrightarrow{H}\left(  x\right)   &  =\frac{1}{2}\int_{\Omega}%
\{\Upsilon_{\alpha_{1}}\left(  x-y\right)  [\frac{\overrightarrow{j}\left(
y\right)  }{\left(  1+k\beta\right)  }-\frac{1}{k}\operatorname*{div}%
\overrightarrow{j}\left(  y\right)  ]\nonumber\\
&  +\Upsilon_{-\alpha_{2}}\left(  x-y\right)  [\frac{\overrightarrow{j}\left(
y\right)  }{\left(  1-k\beta\right)  }+\frac{1}{k}\operatorname*{div}%
\overrightarrow{j}\left(  y\right)  ]\}dy\nonumber\\
&  +\frac{1}{2i}\int_{\Gamma}\{-\Upsilon_{\alpha_{1}}\left(  x-y\right)
\overrightarrow{n}\left(  y\right)  [\overrightarrow{E}\left(  y\right)
+i\overrightarrow{H}\left(  y\right)  ]\nonumber\\
&  +\Upsilon_{-\alpha_{2}}\left(  x-y\right)  \overrightarrow{n}\left(
y\right)  [\overrightarrow{E}\left(  y\right)  -iH\left(  y\right)  ]\}d\Gamma
\end{align}

Las dos \'{u}ltimas igualdades cuaterni\'{o}nicas pueden ser reescritas en
forma vectorial. En este caso las partes escalares no contienen ninguna nueva
informaci\'{o}n ya que representan una variante de la f\'{o}rmula de Gauss,
(vea la explicaci\'{o}n en pag. 120 de \cite{KSbook}) y las partes vectoriales
representan las f\'{o}rmulas de Stratton-Chu (vea por ejemplo \cite{AMS}). En
pocas palabras, estas igualdades cuaterni\'{o}nicas representan resultados
bien conocidos escritos en otra forma.

\bigskip

La consideraci\'{o}n del siguiente problema para el campo electromagn\'{e}tico
nos pone de manifiesto la ventaja del uso del an\'{a}lisis cuaterni\'{o}nico
comparado con la t\'{e}cnica del c\'{a}lculo vectorial.

\bigskip

Problema.- Sea $\Gamma:=\partial\Omega$ una superficie de Liapunov en la cual
est\'{a}n definidos dos vectores $\overrightarrow{e}$ y $\overrightarrow{h}$
que pertenecen al espacio funcional de H\"{o}lder (vea por ejemplo
\cite{Vladimirov}). Encontrar dos vectores $\overrightarrow{E}$ y
$\overrightarrow{H}$ que satisfagan las ecuaciones de Maxwell

\bigskip%

\begin{equation}
\operatorname{rot}\overrightarrow{E}\left(  x\right)  =-ik\left(
\overrightarrow{H}\left(  x\right)  +\beta\operatorname{rot}\overrightarrow
{H}\left(  x\right)  \right)
\end{equation}%

\begin{equation}
\operatorname{rot}\overrightarrow{H}\left(  x\right)  =ik\left(
\overrightarrow{E}\left(  x\right)  +\beta\operatorname{rot}\overrightarrow
{E}\left(  x\right)  \right)
\end{equation}
en $\Omega$ y en la frontera $\Gamma$ coincidan con $\overrightarrow{e}$ y
$\overrightarrow{h}:$

\bigskip%

\[
\overrightarrow{E}\mid_{\Gamma}=\overrightarrow{e},\ \ \overrightarrow{H}%
\mid_{\Gamma}=\overrightarrow{h}
\]

\bigskip

En otras palabras, se requiere extender los vectores $\overrightarrow{e}$ y
$\overrightarrow{h}$ al dominio de $\Omega$ de tal manera que satisfagan las
ecuaciones (44) y (45).

\bigskip

Este problema de extendibilidad para el campo electromagn\'{e}tico es
incorrecto (ill-posed) debido a que no siempre existe una soluci\'{o}n. En el
siguiente teorema damos el criterio de la existencia de la soluci\'{o}n del problema.

Teorema. Dada una superficie cerrada de Liapunov $\Gamma:=\partial
\Omega\subset R^{3}$ y las funciones $\overrightarrow{e},\overrightarrow{h}\in
C^{o,\epsilon}\left(  \Gamma;C^{3}\right)  $, son los valores de frontera de
las soluciones de las ecuaciones \ de Maxwell%

\begin{equation}
\operatorname{rot}\overrightarrow{E}\left(  x\right)  =-ik\left(
\overrightarrow{H}\left(  x\right)  +\beta\operatorname{rot}\overrightarrow
{H}\left(  x\right)  \right)  \label{S57}%
\end{equation}%

\begin{equation}
\operatorname{rot}\overrightarrow{H}\left(  x\right)  =ik\left(
\overrightarrow{E}\left(  x\right)  +\beta\operatorname{rot}\overrightarrow
{E}\left(  x\right)  \right)  \label{S58}%
\end{equation}
en $\Omega$ si y solo si las siguientes igualdades se cumplen:

\bigskip%

\begin{align}
&  \overrightarrow{e}\left(  x\right)  =-\frac{1}{2}\int_{\Gamma}%
\{\Upsilon_{\alpha_{1}}\left(  x-y\right)  \overrightarrow{n}\left(  y\right)
[\overrightarrow{e}\left(  y\right)  +i\overrightarrow{h}\left(  y\right)
]\nonumber\\
&  +\Upsilon_{-\alpha_{2}}\left(  x-y\right)  \overrightarrow{n}\left(
y\right)  [\overrightarrow{e}\left(  y\right)  -i\overrightarrow{h}\left(
y\right)  ]\}d\Gamma
\end{align}%

\begin{align}
&  \overrightarrow{h}\left(  x\right)  =\frac{1}{2i}\int_{\Gamma}%
\{-\Upsilon_{\alpha_{1}}\left(  x-y\right)  \overrightarrow{n}\left(
y\right)  [\overrightarrow{e}\left(  y\right)  +i\overrightarrow{h}\left(
y\right)  ]\nonumber\\
&  +\Upsilon_{-\alpha_{2}}\left(  x-y\right)  \overrightarrow{n}\left(
y\right)  [\overrightarrow{e}\left(  y\right)  -i\overrightarrow{h}\left(
y\right)  ]\}d\Gamma
\end{align}
para cualquier $\tau\in\Gamma$.

\bigskip

Este resultado obviamente puede ser escrito en t\'{e}rminos vectoriales, sin
embargo, su naturaleza es en esencia cuaterni\'{o}nica ya que para obtener el
criterio se necesitan tanto las partes vectoriales de (48) y (49) como
necesariamente las partes escalares de las mismas f\'{o}rmulas. Es un ejemplo
del caso cuando para obtener un resultado en 3 dimensiones hay que hacer uso
de la cuarta dimensi\'{o}n.

\bigskip

Nota: El mismo resultado se puede obtener para el exterior del dominio
$\Omega$.

\end{document}